\documentclass[galley,grl]{agu2001}
\usepackage{dcolumn}% Align table columns on decimal point
\usepackage{bm}% bold math

\usepackage{lineno}
\usepackage{graphicx}

\authorrunninghead{URITSKY ET AL.}
\titlerunninghead{SCALING REGIMES IN EARTH'S MAGNETOSPHERE}
\authoraddr{V. M. Uritsky, Physics and Astronomy Department, University of Calgary,
SB605, 2500 University Drive NW, Calgary, AB T3A0P4, Canada. (vuritsky@phas.ucalgary.ca)}

\linenumbers*[1]

\begin{document}

%% ------------------------------------------------------------------------ %%
%  ENABLE IMAGE DISPLAY WHILE USING DRAFT MODE
%% ------------------------------------------------------------------------ %%
% Uncomment the following code (as well as \usepackage{graphicx} above)
% if you need to include images in draft mode
\setkeys{Gin}{draft=false}
% PLEASE NOTE: WHEN YOU SUBMIT YOUR LATEX FILE TO GEMS, COMMENT OUT ANY COMMANDS
% THAT INCLUDE GRAPHICS.

\title{Distinct Scaling Regimes of Energy Release Dynamics in the Nighttime Magnetosphere}

%\author{V. M. Uritsky, E. Donovan, E. Spanswick}
%\affil{Physics and Astronomy Department,
%University of Calgary, Calgary, AB, Canada}

%\author{A. J. Klimas}
%\affil{UMBC at NASA / Goddard Space Flight Center, Greenbelt, Maryland, USA}

\author{V. M. Uritsky \altaffilmark{1}, E. Donovan \altaffilmark{1},
A. J. Klimas \altaffilmark{2}, and E. Spanswick \altaffilmark{1}}

\altaffiltext{1}
{Physics and Astronomy Department, University of Calgary, Calgary, AB, Canada}

\altaffiltext{2}
{UMBC at NASA / Goddard Space Flight Center, Greenbelt, Maryland, USA}

\begin{abstract}

Based on a spatiotemporal analysis of POLAR UVI images, we show that the auroral emission events that initiate equatorward of the isotropic boundary (IB) obtained from a time-dependent empirical model, have systematically steeper power-law slopes of energy, power, area and lifetime probability distributions compared to the events that initiate poleward of the IB. The low-latitude group of  events contains a distinct subpopulation of substorm-scale disturbances violating the power-law behavior, while the high latitude group is described by nearly perfect power-law statistics over the entire range of scales studied. The results obtained indicate that the inner and outer portions of the plasma sheet are characterized by substantially different scaling regimes of bursty energy dissipation suggestive of different physics in these regions. 

\end{abstract}

\begin{article}
% If using draft mode \end{article} must follow the references section.

\section{Introduction}
%\subsection{Level 2 Head} An example.

The activity of the nighttime auroral oval represents a wide range of dynamical processes in the magnetotail, including substorm expansion onsets, pseudobreakups, steady magnetospheric convection events with or without substorms, bursty bulk flows, and sawtooth events (see e.g., \cite{zesta00, lui01, frey04, henderson06}). Despite the diversity of physical conditions associated with each particular type of auroral activity, their net energy output can be described by a set of apparently universal power-laws (\cite{lui00, lui02,uritsky03, uritsky02, uritsky06}) signaling the existence of a organizing dynamical principle arranging intermittent magnetospheric dissipation across vast ranges of spatial and temporal scales.

Power-law intermittency of energy dissipation has attracted significant attention in modern statistical mechanics (see \cite{dhar06} and refs therein) and is often considered a hallmark of turbulent and/or critical phenomena with no characteristic scales other than those dictated by the finite size of the system (\cite{sreenivasan04, lubeck04}). Examples of such behavior in geo- and space sciences include fully developed turbulence in hydrodynamic or magnetized flows (\cite{lazarian06}), Guttenberg-Richter statistics of earthquake magnitudes (\cite{turcotte89}), scale-invariance in the solar corona (\cite{charbonneau01}). In this context, the auroral activity provides one of the most impressive examples of scale-free behavior in nature. The energy distribution of electron emission regions exhibits a power-law shape over a range of 6 orders of magnitude (\cite{uritsky02}) which can be extended to up to 11 orders by combining the satellite data with ground-based TV observations (\cite{kozelov04}). 

The auroral emission statistics reported so far represent global long-term properties of nighttime magnetospheric disturbances. The fact that these properties are dominated by power-law scaling does not eliminate the possibility of a more complex behavior on the level of specific plasma sheet structures described by drastically different physical conditions and geometry. 
%Exploring these effects is necessary for developing more solid theoretical links between the statistical and the dynamical plasma descriptions, evaluating predictability of various classes of geomagnetic disturbances, and obtaining occurrence frequency estimates of various types of activity for planning future missions.
In this study, we are taking a step toward a better understanding of the relationship between the scale-free auroral precipitation statistics and the underlying central plasma sheet (CPS) morphology. We suggest that the inner and the outer CPS regions are responsible for three distinct scaling modes of the auroral precipitation dynamics, and provide a possible physical interpretation for the observed differences.
%We find that activity originating in the inner CPS is characterized by substantially higher exponents for energy, area, power, and lifetime power-law probability distributions compared to the corresponding distributions for activity initiated in the outer CPS region.
%In addition, the inner CPS events exhibit a crossover separating what we believe are substorm energy scales from small-scale activity, while the tail CPS is a source of more robust scale-free auroral dynamics described by remarkably stable power-law distributions. The emission events produced in the outer CPS have significantly higher relative occurrence compared to the inner CPS events and therefore dominate the overall auroral statistics.

\section{Data and Algorithm}

We have studied time series of digital images of nighttime northern aurora (55-80 MLat, 2000 - 0400 MLT) taken by the Ultraviolet Imager (UVI) onboard the POLAR spacecraft in the 165.5 to 174.5 nm portion of the Lyman-Birge-Hopfield spectral band (integration time 36.5 s, time resolution 184 s).  The data analyzed include 16,000 images covering two observation periods: 01/01/1997 - 02/28/1997 and 01/01/1998 - 02/28/1998. Our analysis was based on spatiotemporal tracking of auroral emission events (\cite{uritsky02, uritsky03}). The UV luminosity $w(t,\bm{r})$ was studied as a function of time $t$ and position $\bm{r}$ on the image plane. First, active auroral regions were identified by applying an activity threshold $w_a$ representing a background UV flux. Contiguous spatial regions with $w(\bm{r},t) > w_a$ were treated as pieces of evolving events. Second, by checking for overlap of common pixels between each pair of consecutive UVI frames, we constructed a set of spatiotemporal integration domains $\Lambda_i (i=1,..,N)$ corresponding to each of $N$ individual emission events found by our method. These domains of contiguous activity in space and time were used to compute the lifetime, $T_i$, the energy,
$E_i=k \int_{\Lambda_i} w(\bm{r},t) \, d\bm{r}dt$, the peak power, $W_i=k \max\limits_{t} (\int_{\Lambda_i(t)} w(\bm{r},t)\, d\bm{r} )$,
as well the peak area, $A_i=\max\limits_{t} (\int_{\Lambda_i(t)} \, d\bm{r} )$ of every event, where $k=2.74\times10^{-8}$ J $\cdot$ photon$^{-1}$ (\cite{brittnacher97}) and the integrals were numerically approximated by sums. The statistics reported below are for the threshold $w_a$=10 photons $\cdot$ cm$^2$ $\cdot$ s$^{-1}$. Their main features remain the same if the threshold is varied at least within the range 5 to 15 photons $\cdot$ cm$^2$ $\cdot$ s$^{-1}$.

The auroral onset positions of each event were estimated with an error of about 300 km in either spatial direction. We organized the data to allow us to compare events that likely originate in the inner magnetosphere relative to events that likely originate in the outer magnetosphere. To accomplish this, we determine the event locations relative to the isotropic boundary (IB) at the meridian of the event origin \footnote{A more accurate analysis based on Tsyganenko T96 and T05 field-line models as well as optical determination of the isotropic boundary will be published elsewhere.}: 
\begin{aguleftmath}
\phi_i = MLat_i - \left[ A_0 - A_1 \ cos \left(\pi(MLT_i - MLT_0)/12 \right) \right]
\label{eq1}
\end{aguleftmath}
Here, $MLT_i$ and $MLat_i$ are the onset coordinates of the $i$-th event. $A_0$, $A_1$, $MLT_0$ are the coefficients of the empirical IB model due to \cite{gvozdevsky95} which were computed based on hourly values of auroral electrojet index (WDC for Geomagnetism, Kyoto) and solar wind dynamic pressure (ACE spacecraft) at the beginning of each auroral intensification. The ions maintain their isotropy on the poleward side of IB due to the effective pitch angle scattering in the tail current sheet (\cite{newell96}). As a result, the latitude of IB has a high correlation with the magnetic field inclination at the geomagnetic equator as well as with the equatorward boundary of the proton aurora as determined from direct optical observations (\cite{donovan03}). Here, we have used the empirical IB model to separate inner and outer emission events. Positive values of $\phi$ on average indicate events originating on field lines that map further out on the plasma sheet, as opposed to negative values corresponding to the events that originate closer to Earth.

The statistics of the emission events that initiated on the poleward ($\phi > 0$) and on the equatorward ($\phi <0$) side of the IB are characterized by sets of probability density distributions $p(x)$, where $x \in \{E, W, A, T \}$. The power-law exponents obtained from these distributions are denoted as $\tau_x$, with the subscript indicating the variable under study.

\section{Results and Discussion}

In the end we have 7481 events, 6231 and 1250 poleward and equatorward of the IB, respectively. These yielded a database of parameters $E$, $W$, $A$, $T$, $MLT$, $MLat$, $\phi$ for those events. Figure 1 is the emission energy $E$ as a function of the relative onset latitude $\phi$ for all events ($n=7481$). The scatterplot has clear asymmetric shape suggesting different statistics for negative and positive $\phi$ values. As we discuss below, these statistical subsets are characterized by significantly different regimes of scaling behavior indicating different physical environments.

Figure 2 shows probability distributions for the events which initiated on the poleward and equatorward sides of the IB. The high-latitude (HL) events ($\phi>0$) exhibit stable power-law distributions of emission energy $E$, peak emission power $W$, lifetime $T$ and peak area $A$ (not shown) with near-zero regression errors. The ranges of scales of these power-law behaviors involves both small auroral activations and large events whose energy output lies in the range of global substorms. The exponents $\tau_E$, $\tau_W$, $\tau_A$ and $\tau_T$ are close to the values obtained earlier for the same observation period without filtering the activity by the onset location (\cite{uritsky02}). 

The low-latitude (LL) events ($\phi<0$) have more complicated statistical properties as can be expected from Fig.~1. Their distribution functions demonstrate a crossover behavior involving small-scale regions described by $\tau_x$ exponents which are greater than the corresponding exponents of HL events, as well as large-scale regions where the slopes are significantly shallower (the $\tau_x$ exponents for these regions are not well defined due to the insufficient number of large-scale LL events). The position of the crossover $E^*=5\times10^{12}$J on the $p(E)$ distribution of LL events matches the data gap in Fig.~1 separating low- and high-energy events with negative relative latitudes $\phi$.

As can be seen from Table 1, only 1/6 of the events contribute to the population with $\phi < 0$. However, their participation in the energy budget of the electron aurora approaches 50 percent as can be seen from the comparison of the overall energy $E_{tot}$ released by LL and HL events. The fact that the HL events are the the most common form of emission dynamics (relative occurrence frequency 82.3$\%$) is not a surprise as these events likely map to the reconnection regions of the plasma sheet, as we briefly discuss below. The table also shows that the scaling behaviors of LL and LH events are described by subsantially different sets of $\tau$ exponents.

Table 2 provides scaling exponents of energy distributions $p(E)$ as well as several parameters reflecting the state of the solar wind and the auroral magnetosphere at the beginnings of emission events for four non-overlapping ranges of magnetic latitude $MLat$. The data show that the group of events that initiated southward of MLat=65 are characterized by significantly higher statistical values of solar wind dynamic pressure and electrojet index. These events seem to appear in a rather stretched magnetotail configuration as reflected by the magnetotail (MT) index (\cite{gvozdevsky95}). One can also notice that the lower the magnetic latitude, the higher is the exponent $\tau_E$. This tendency is in accordance with the difference in the LL- and HL-event statistics reported in Table 1, and it also suggests that the change of the scaling regime of auroral emission events with onset latitude is a continuous process rather than a discrete transition fully dictated by the empirical IB location.

From the statistical mechanics viewpoint, the existence of two or more scaling regimes within the same physical system signals the presence of several distinct universality classes (UC) in the underlying turbulent dynamics (\cite{lubeck04, dhar06}). It is known that many stochastic scale-invariant phenomena can be mapped onto a finite collection of such classes. This mapping is insensitive to a variety of physical parameters and conditions and can be the same for quite different systems. Typically, each UC brings its own set of critical exponents which express its inherent set of relevant variables and the underlying small- and large-scale symmetries (\cite{benhur96}). In this context, our results clearly show that the energy release dynamics in the inner and outer CPS belong to different UCs and are therefore governed by different physical mechanisms.

The nature of the isotropic boundary sheds some light on the scaling regimes observed. On average, HL events tend to initiate in the outer CPS region. The energy conversion in this region is believed to be dominated by magnetic reconnection ({\cite{birn81}}). Our results strongly suggest that near-Earth (midtail) reconnection is a turbulent bursty process with no well-defined dissipation scales. Judging by the relative occurrence of these events (Table 1), the scale-free activity in the reconnection region is the dominant mode of energy release dynamics in the nightside  magnetosphere. The power-law emission exponents obtained for this region are rather close to the exponents from a driven current sheet simulation (\cite{klimas04}) consistent with multiscale turbulent reconnection being the source of scale-invariance in the outer CPS dynamics. The LL events are mainly produced in the inner CPS. With a less stretched magnetic field topology, this region is not a region that favors magnetic reconnection; however, it can be prone to current disruption which offers an alternate mechanism for energy release in the inner tail (\cite{lui00cd}). 

Our findings can be summarized as follows: (1) the high-latitude emission events are characterized by broad-band power-law statistics with no characteristic scales; (2) the low-latitude events constitute a non-uniform statistical population with an energy crossover separating large and small scale activity; (3) the scaling exponents of the low- and high-latitude events are significantly different within the entire range of scales studied, which signals distinct universality classes of energy release dynamics. To the extent that the empirical IB model can be considered an accurate representation of the true IB, the onsets of the HL events are located in an outer CPS region, the onsets of the LL events in an inner CPS regions, and these two regions exhibit distinctly different energy release dynamics.

The difference in UC following from this analysis is suggestive of different physics. It would be a significant step forward if we could use physics-based models of CPS dynamics to predict UCs, and to thereby determine how the different statistical behaviors of the inner and outer regions might be generated.

%Future possible research avenues that would follow on to this study include the following. 

%The difference in UC following from this analysis is suggestive of different physics. As we are studying dynamic events that include substorms, pseudobreakups and other forms of high-latitude activations, this difference may be as straightforward as current disruption versus magnetic reconnection scenarios of the CPS response, or something more subtle. This difference can also responsible for the two classes of earthward fast flows recently reported by \cite{shue08}. It would be a significant step forward if we could use physics-based models of CPS dynamics to predict UCs, to determine how the different statistical behaviors of the inner and outer regions might be generated.

\begin{acknowledgments}
This work was partly supported by NSERC operating grant of EFD. We thank William Liu for stimulating discussions.
\end{acknowledgments}

%FIGURES AND TABLES

%----------------------------------  FIG 1 ------------------------------------------
\begin{figure}[htbp]
\hskip -0.3cm
\noindent\includegraphics*[width=8.5 cm]{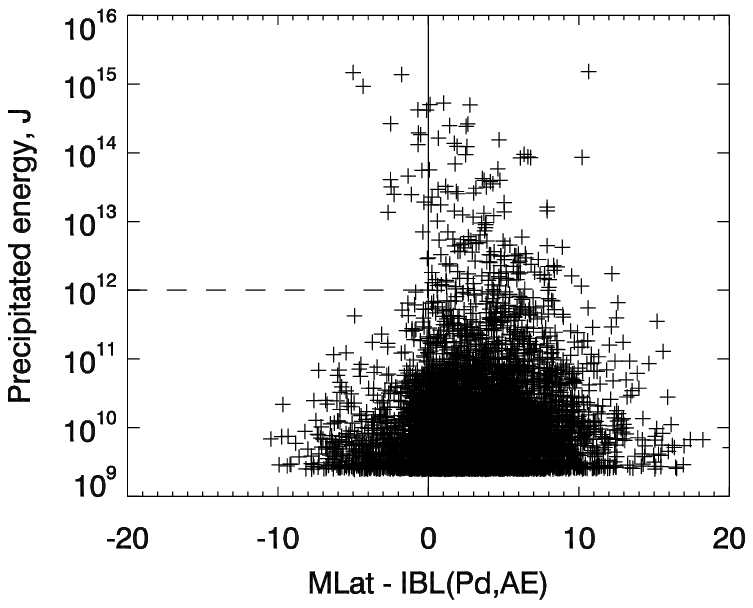}

\caption{\label{Fig1} Scatterplot of emission energies $E$ versus relative magnetic latitudes $\phi$ of onset locations measured with respect to the empirical IB latitude (IBL) (see eq.(\ref{eq1})). The high-latitude ($\phi>0$) and the low-latitude ($\phi<0$) events contribute to two distinct scaling regimes of energy release dynamics in the nighttime magnetosphere as discussed further in the text. The dashed horizontal line marks the position of the crossover on the energy distribution shown in Fig.~2. The sharp lower cutoff is due to the fixed activity threshold $w_a$.}
%\vskip -0.3cm
\end{figure}
%----------------------------------  FIG 2 ------------------------------------------
\begin{figure}[htbp]
%\noindent\includegraphics*[width=4.2cm]{fig2a}
%\hskip -0.3cm
%\noindent\includegraphics*[width=4.2cm]{fig2b}
%\noindent\includegraphics*[width=4.2cm]{fig2c}
%\hskip -0.3cm
%\noindent\includegraphics*[width=4.2cm]{fig2d}
%\caption{\label{fig2} Probability density distributions (top) and cumulative distributions (bottom) of auroral emission energy $E$ and peak emission power $W$ constructed for the events initiated poleward ($\phi>0$) and equatorward ($\phi<0$) of IB. The low-latitude distributions are shifted for easier comparison. The dotted lines show the log-log distribution slopes for small-scale low-latitude emissions events (mode $A_1$) and for the entire group of high-latitude events (mode $B$).}

\noindent\includegraphics*[width=7.5 cm]{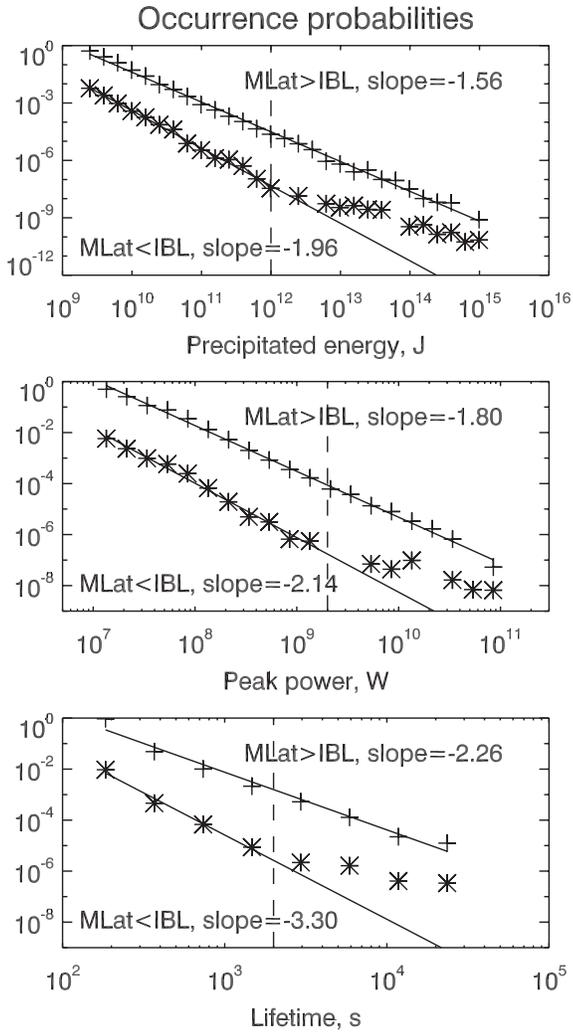}

\caption{\label{fig2} Probability distributions of energy $E$, peak power $W$, and lifetime $T$ of auroral emission events initiated poleward ($\phi>0$, crosses) and equatorward ($\phi<0$, stars) of the empirical IB. The low-latitude distributions are shifted for easier comparison. The solid lines show the log-log distribution slopes for small-scale LL events (the power-law portions to the left of the dashed vertical lines) and for the entire group of HL events.}
\end{figure}
%----------------------------------  FIG 3 ------------------------------------------

\begin{table}
\caption{\label{tab:table1} Comparative characteristics of low- and high-latitude emission events.}
%\begin{ruledtabular}

\begin{tabular}{lcc}
\hline
Type of events 		  		&  LL ($\phi<0$)          &   HL  ($\phi>0$)           \\
\hline
\\
$\#$ of events 					& 1250 (16.7$\%$)            & 6231 (82.3$\%$) \\
$E_{tot}$,$J$   			 &	$5.7\times10^{15}$ (47.5$\%$)         & $6.3\times10^{15}$ (52.5$\%$)  \\
\\
$\tau_E$                & $1.96\pm0.04$    & $1.56\pm0.02$      \\

$\tau_W$                & $2.14\pm0.07$    & $1.80\pm0.02$      \\

$\tau_T$                & $3.30\pm0.26$    & $2.26\pm0.13$      \\

$\tau_A$                & $2.30\pm0.10$    & $1.93\pm0.06$      \\

%\multicolumn{4}{c}{{\it Suggested interpretation}}\\
%Location								& \ \ \ \ \ \ \ \ \  \ \ \ \ \ \ Inner				& $ \!\!\!\!\!\!\!\!\!\!\!\!\!\!\!\!\!\!\! $ CPS 				&	Outer CPS \\	
%Mechanism					      & \ \ \ \ \ \ \ \ \  Current 		& $ \!\!\!\!\!\!\!\!\!\!\!\!\!\!\!\!\!\! $ disruption 	& Reconnection	\\
\\
\hline   						          
\end{tabular}
%\end{ruledtabular}
\end{table}
%========================================================================

\begin{table}
\caption{\label{tab:table2} Scaling exponent $\tau_E$ of emission events, solar wind dynamic pressure ($P_d$, auroral electrojet ($AE$) index and the magnetotail ($MT$) index (IB latitude at $MLT$=0, (eq. \ref{eq1})), for four ranges of onset  $MLat$.} 
%\begin{ruledtabular}

\begin{tabular}{lcccc}
\hline
\\
$MLat$ 		& $\tau_E$      & $Pd$, nPa      & $AE$, nT    &   $MT$         \\
\hline
\\
$ < 65$				& $2.06\pm0.04$ & $2.64\pm0.05$  & $ 206\pm5 $ & $ 63.2\pm0.03 $ \\
$65...67$  & $1.88\pm0.04$ & $2.42\pm0.03$  & $ 106\pm3 $ & $ 63.9\pm0.03 $ \\
$67...69$  & $1.59\pm0.03$ & $2.41\pm0.03$  & $  84\pm3 $ & $ 64.0\pm0.03 $ \\
$>69$       & $1.58\pm0.03$ & $2.20\pm0.03$  & $  93\pm4 $ & $ 64.1\pm0.03 $ \\
\\
\hline   						          
\end{tabular}
%\end{ruledtabular}
\end{table}

\bibliographystyle{agu04}
%\bibliography{grl_2007} % Produces the bibliography via BibTeX. 

% To continue usingg BibTeX:
% 1. Remove the comment mark from  `` %\bibliography{grl_2007}'' above
% 2. REMOVE everything below between % +++++ and % +++++

% +++++

% +++++

\end{article}

\end{document}